\documentstyle[prd,aps]{revtex}
\begin{document}
\large

\title{\Large\bf An anisotropic cosmological model with isotropic
background radiation}
\author{\large Saulo Carneiro$^{1,2}$ and Guillermo A. Mena Marug\'{a}n$^1$
\vspace*{.3cm}}
\address{\large {\it $^1$ Centro de F\'{\i}sica Miguel A. Catal\'{a}n, IMAFF, CSIC,
Serrano 121, 28006 Madrid, Spain\\$^2$ Instituto de F\'{\i}sica,
Universidade Federal da Bahia, 40210-340, Salvador, Bahia,
Brazil}\vspace*{.5cm}} \maketitle

\begin{abstract}\large{\bf Abstract.}
We present an exact solution of Einstein equations
that describes a Bianchi type III spacetime with conformal
expansion. The matter content is given by an anisotropic scalar
field and two perfect fluids representing dust and isotropic
radiation. Based on this solution, we construct a cosmological
model that respects the evolution of the scale factor predicted in
standard cosmology.
\end{abstract}
\vspace*{.7cm}

A crucial question in cosmology is whether the observed isotropy
of the cosmic microwave background (CMB), together with the
apparent homogeneity and isotropy of clustering matter, suffices
to guarantee the Cosmological Principle. An affirmative answer is
strongly supported by a theorem proved by Ehlers, Geren and Sachs
\cite{EGS}. The EGS theorem ensures that a spacetime region
satisfying Einstein equations and containing only dust and
radiation has a Friedmann-Robertson-Walker (FRW) geometry provided
that the dust velocity field $u^{\mu}$ is (geodesic and)
expanding, and that the distribution function of the photons is a
solution to the Liouville equation which is isotropic with respect
to $u^{\mu}$. In fact, the result that the geometry is FRW depends
critically on the hypotheses of the theorem, and there exist
counterexamples in which all but one of the assumptions are
satisfied \cite{FMP,CE,CB}. The theorem admits generalizations for
matter consisting in a generic (barotropic) perfect fluid
\cite{FMP}, and for an almost isotropic CMB \cite{SME}, case in
which the geometry is approximately of the FRW form.

We will concentrate our discussion on the validity of the EGS
theorem when one relaxes the assumption about the kind of matter
content. This is important in modern cosmology because, in order
to explain the observed acceleration detected with type Ia
supernovae, scalar fields are often introduced to generate a
quintessence component. It was recently shown that the conclusions
of the EGS theorem remain valid in the presence of a quitessence
field unless the gradient of the field is orthogonal to the dust
congruence \cite{CB}. This alternative to the FRW geometry has
been neglected so far because it is believed to lead to unphysical
situations. We will show, nevertheless, that it is possible to
find physically acceptable systems in which the commented
alternative allows one to circumvent the Cosmological Principle.

The most general line-element and Einstein tensor compatible with
an isotropic radiation can be found in \cite{FMP}. The aim of the
present work is to explicitly construct a feasible solution to the
Einstein equations respecting this isotropy of the CMB. This task
includes determining not only the metric, but also the matter
content. Furthermore, we want such a content to be of cosmological
and physical interest, in the sense that it must be composed of
dust, radiation, and a scalar field, and that none of the energy
conditions (weak, strong or dominant) is violated. In addition, we
will accept that both the perfect fluid components and the spatial
sections of the spacetime are homogeneous.

Our ansatz for the metric is
\begin{equation}
\label{met} ds^2=a^2(\eta)\left[-d\eta^2+
dx^2+e^{2x}dy^2+dz^2\right].
\end{equation}
This is a Bianchi type III metric, with Killing vectors given by
$\partial_x-y\partial_y$, $\partial_y$, and $\partial_z$. It can
be considered as the vanishing-rotation member of a family of
line-elements with application in rotating cosmology \cite{KO}
that are expanding versions of a class of G\"{o}del-like metrics
\cite{RT}. The spatial topology is the direct product of a
pseudosphere and a real line \cite{RT}. The velocity field
$u^{\mu}=\delta^{\mu}_{\eta}/a$ is geodesic, shear-free and
vorticity-free. Its expansion is $3a^{\prime}/a^2$, where the
prime stands for the derivative with respect to the conformal time
$\eta$. Besides, metric (\ref{met}) possesses a conformal Killing
vector (CKV) $\xi^{\mu}=a(\eta) u^{\mu}$. Actually, the existence
of this CKV proportional to the velocity $u^{\mu}$ suffices to
guarantee the isotropy of the CMB (for comoving radiation), as
well as the absence of parallax effects \cite{KO}.

The Einstein equations for this metric are
\begin{equation}
\label{Eeq} Ea^4=3(a^{\prime})^2-a^2,\hspace*{.8cm}P_xa^4 =
P_ya^4=P_za^4-1=(a^{\prime})^2-2aa^{\prime\prime},
\end{equation}
where $E\equiv -T^{\eta}_{\eta}$ is the energy density and
$P_i\equiv T^i_i$ ($i=x,y$, or $z$) are the principal pressures.
As for the matter content, let us start by introducing a massless,
minimally coupled scalar field $\Phi$ with vanishing
self-interaction potential. The dynamical equation for such a
field is $\Phi_{;\mu\nu}g^{\mu\nu}=0$, with $g^{\mu\nu}$ being the
inverse of the metric and the semicolon denoting the covariant
derivative. A solution is $\Phi=Cz$, where $C$ is any constant
number. It is worth remarking that, on this solution, the gradient
of $\Phi$ is actually orthogonal to the four-velocity $u^{\mu}$.
Besides, the energy-momentum tensor of the scalar field
$T^{\mu}_{\nu}=\Phi_{,\nu}\Phi^{,\mu}-\Phi_{,\sigma}\Phi^{,\sigma}
\delta^{\mu}_{\nu}/2$ becomes diagonal, with
$E^{(\Phi)}=P_z^{(\Phi)}=-P_x^{(\Phi)}=-P_y^{(\Phi)}=C^2/(2a^2)$.
It is straightforward to check that all of the energy conditions
are then satisfied, so that the solution is physically viable.
Moreover, choosing $C=1$ (or $C=-1$ with a flip of sign in $z$),
one can in fact absorb in the contribution of the scalar field all
the anisotropic pressures appearing in (\ref{Eeq}).

Therefore, accepting the solution $\Phi=z$ and denoting by
$\varrho=E-E^{(\Phi)}$ and $p=P_i-P_i^{(\Phi)}$ the energy density
and (isotropic) pressure of the remaining matter components, we
conclude that this additional matter must satisfy the Einstein
equations
\begin{equation}
\label{Neq} \varrho a^4=
3(a^{\prime})^2-\frac{3}{2}a^2,\hspace*{.8cm} pa^4=
(a^{\prime})^2-2aa^{\prime\prime}+\frac{a^2}{2},
\end{equation}
which are precisely those corresponding to a perfect fluid in open
FRW cosmology, except for changes of scale by factors of order
unity:
\begin{equation}
\eta=\sqrt{2}\eta_{F},\hspace*{.8cm}
\sqrt{2}\;a\!\left(\eta=\sqrt{2}\eta_{F}\right)=a_{F}(\eta_{F}).
\end{equation}
Here, $\eta_F$ and $a_F$ are the conformal time and scale factor
of the standard open FRW model.

We are now in an adequate position to obtain the solution that we
were seeking. We assume that the remaining matter consists of dust
and radiation, which form a two-component perfect fluid with
four-velocity $u^{\mu}$. The energy density and pressure are thus
given by $\varrho=A^2/a^4+D/a^3$ and $p=A^2/(3a^4)$, where $A$ and
$D$ are two positive constants. One then arrives at the following
solution of (\ref{Neq}):
\begin{equation}
\label{sol}
a=\frac{D}{3}\left[\cosh\left(\frac{\eta}{\sqrt{2}}\right)-1\right]+
\sqrt{\frac{2}{3}}A\sinh\left(\frac{\eta}{\sqrt{2}}\right).
\end{equation}
Inverting this formula and integrating $dt=ad\eta$, one gets the
time expressions
\begin{eqnarray}
\label{confo} \eta &=& \sqrt{2}
\ln\left[\frac{3a+D+\sqrt{9a^2+6Da+6A^2}}{D+\sqrt{6}A}\; \right],
\\ \label{t}
t&=&\frac{D}{3}\left[-\eta+\sqrt{2}\sinh\left(\frac{\eta}{\sqrt{2}}
\right)\right]+\frac{2A}{\sqrt{3}}
\left[\cosh\left(\frac{\eta}{\sqrt{2}}\right)-1\right].
\end{eqnarray}

Clearly, this solution leads to a cosmological model that
coincides in most of its predictions with those of an open FRW
scenario. In particular, the thermal history and the cosmological
parameters that depend only on the evolution of the scale factor
are essentially the same as in standard (open) FRW cosmology. For
instance, for the Hubble parameter $H$ and the deceleration
parameter $q$ one obtains
\begin{equation} \label{H}
H=\frac{a^{\prime}}{a^2} = \sqrt{ \frac{3a^2+2Da+
2A^2}{6a^4}},\hspace*{.6cm} q=1
-\frac{aa^{\prime\prime}}{(a^{\prime})^2}=
\frac{Da+2A^2}{3a^2+2Da+2A^2}>0.
\end{equation}
Note that the expansion is not accelerated, so that (at least in
this sense) the model cannot be considered fully realistic.
Nonetheless, this problem can be overcome as in FRW cosmology,
i.e., with the introduction of a cosmological constant or an
additional scalar field with accelerating properties.

In spite of the similarities of the model with an open FRW
universe, there exist some differences between the two
cosmological scenarios. For instance, the standard epochs of
radiation and dust domination are now followed by a scalar-field
dominated era. During these eras, the scale factor ranges in $0<
a\leq A^2/D$ (radiation), $A^2/D\leq a\leq 2D$ (dust), and $2D\leq
a$ (scalar field). We suppose that $A\ll D$, so that there exists
a large epoch dominated by clustering matter. A more relevant
discrepancy arises in the relative energy density. In our case,
this density includes also the contribution of the scalar field.
Using (\ref{H}), we get
\begin{equation}\label{O} {\mathrm \Omega}= \frac{E}{3H^2} =
\frac{a^2+2Da+2A^2}{3a^2+ 2Da+2A^2}.\end{equation} One can check
that ${\mathrm \Omega}$ decreases with the expansion. At the
big-bang, ${\mathrm \Omega}$ equals the unity, like in FRW
cosmology. However, when the scale factor expands to infinity,
${\mathrm \Omega}$ does not vanish, but tends to one third.
Therefore, the contribution of the scalar field ensures that the
energy density is of the order of the critical one during the
whole evolution of the universe, leading to a quasi-flat
cosmology.

To estimate the constants $A$ and $D$ that appear in our model,
the parameters $q$ and ${\mathrm \Omega}$, and the current values
of $a$ and $t$, we can proceed as follows. From (\ref{H}),
\begin{equation}
a_0=\frac{1}{H_0}\sqrt{\frac{1+z_{{\mathrm e}{\mathrm
q}}}{2\left(1-{\mathrm \Omega}_0^{({\mathrm
d})}\right)(1+z_{{\mathrm e}{\mathrm q}}) -2{\mathrm
\Omega}_0^{({\mathrm d})}}}\;,\end{equation} where the subscript
$0$ stands for evaluation at present and $z_{{\mathrm e}{\mathrm
q}}=-1+{\mathrm \Omega}_0^{({\mathrm d})}/{\mathrm
\Omega}_0^{({\mathrm r})}$ is the redshift at equilibrium between
the contributions to ${\mathrm \Omega}$ of dust and radiation,
denoted by ${\mathrm \Omega}^{({\mathrm d})}$ and ${\mathrm
\Omega}^{({\mathrm r})}$. In addition, $D=3H_0^2a_0^3{\mathrm
\Omega}_0^{({\mathrm d})}$ and $A^2=Da_0/(1+z_{{\mathrm e}{\mathrm
q}})$. Using the values $H_0=65$ km/(sMp), ${\mathrm
\Omega}_0^{({\mathrm d})}=0.35$, and $z_{{\mathrm e}{\mathrm
q}}=5000$, we get $A=1.6\times10^{24}\;{\mathrm m}$,
$D=1.0\times10^{26}\; {\mathrm m}$, $a_0=1.2\times10^{26}\;
{\mathrm m}$, $t_0=12$ Gyr, $q_0=0.18$, and ${\mathrm
\Omega}_0=0.57$. Note that $A$ is really much smaller than $D$, as
we had assumed. Since $a_0$ is smaller but close to $2D$, the
universe would be at the end of the dust-dominated era. In
addition, we notice that the age of the universe $t_0$, although
very close, is still beyond the lower bounds obtained from
observation.

Finally, it is worth commenting that, though the redshift of the
radiation emitted by dust particles depends only on the emission
and reception times, the fact that the metric is anisotropic
implies that the distance to astrophysical objects with identical
redshift varies with the direction of observation. However, it is
possible to show that these anisotropies in the measurements of
distances are not large enough to conflict with observational data
\cite{CM}. Actually, one can prove that these anisotropies
increase with the redshift and that, with the values estimated
above for the parameters of our model, the maximum variation in
the angular diameter distance of a source at unit redshift is less
than five per cent, which is not currently detectable \cite{CM}.

In conclusion, we have shown that, introducing a scalar field, one
can find anisotropic solutions of the Einstein equations that are
compatible with the expansion of the universe, the isotropy of the
CMB, and the homogeneity and isotropy of clustering matter which
follows geodesics. We have proved this fact by explicitly
constructing a solution with the required properties. Such a
solution leads to a cosmological model that possesses most of the
desirable features of a standard FRW scenario.

\end{document}